\pdfoutput=1
\documentclass[10pt,a4paper,english,aps,showpacs]{revtex4}
\usepackage[T1]{fontenc}
\usepackage[latin9]{inputenc}
\usepackage{color}
\usepackage{babel}
\usepackage{float}
\usepackage{booktabs}
\usepackage{amsmath}
\usepackage{graphicx}
\usepackage{amssymb}

\makeatletter

\providecommand{\tabularnewline}{\\}

\makeatother

\begin{document}

\title{Angular-planar CMB power spectrum}

\author{Thiago S. Pereira}

\email{thiago@ift.unesp.br}

\affiliation{Instituto de F\'isica Te\'orica, UNESP - Universidade Estadual Paulista, Caixa Postal 70532-2, 01156-970, S\~ao Paulo, Brasil}

\author{L. Raul Abramo}

\email{abramo@fma.if.usp.br}

\affiliation{Instituto de F\'isica, Universidade de S\~ao Paulo, CP 66318, 05314-970,
S\~ao Paulo, Brasil}
\begin{abstract}
Gaussianity and statistical isotropy of the universe are
modern cosmology's minimal set of hypotheses.
In this work we introduce a new statistical test to
detect observational deviations from this minimal set. By defining the
temperature correlation function over the whole celestial
sphere, we are able to independently quantify both \textit{angular} and
\textit{planar} dependence (modulations) of the CMB temperature power spectrum over
different slices of this sphere. Given that planar dependence leads to further modulations 
of the usual angular power spectrum $C_{\ell}$, this test can potentially reveal richer 
structures in the morphology of the primordial temperature field. We have also 
constructed an unbiased estimator for this angular-planar power spectrum which naturally 
generalizes the estimator for the usual $C_\ell$'s. With the help of a chi-square analysis, we have used this estimator to search for observational deviations of statistical isotropy in 
WMAP's 5 year release data set (ILC5), where we found only slight anomalies
on the angular scales $\ell=7$ and $\ell=8$. Since this angular-planar statistic is model-independent, it is ideal to employ in searches of statistical anisotropy (e.g., contaminations from the galactic plane) and to characterize non-gaussianities.
\end{abstract}


\maketitle

\section{introduction\label{intro}}

Many efforts have been made towards understanding the statistical
properties of the cosmic microwave background (CMB) temperature field
in the past few years. The main motivation behind these efforts is
that, in a homogeneous and isotropic universe in which inflation is
driven by a single canonical scalar field, the primordial temperature
field is set by Gaussian and statistically isotropic physical processes.
Since nonlinear evolution destroys all these putative initial gaussianities,
we must search for any fundamental deviations from these statistical
properties at early epochs
and as close as possible to the linear regime. This makes the CMB
the ideal physical observable to employ in searches of statistical anisotropies
and non-gaussianities. Any significant observational deviation from
this picture could reveal us something as yet unsuspected about the
basic nature of our universe.

While this program seems to be well motivated by itself, careful analysis
of recent temperature maps obtained by the WMAP team \citep{Hinshaw:2008kr,Komatsu:2008hk,Nolta:2008ih}
have hinted at some apparent anomalies -- mainly
in the low multipoles sector \citep{deOliveiraCosta:2003pu,Eriksen:2003db,Schwarz:2004gk,Land:2005ad,Eriksen:2007pc}.
If one leaves aside for a moment the perennial problem of \textit{a posteriori} 
statistics \citep{Abramo:2006gw}, these findings raise the possibility that
the anomalies could be a first hint towards some new physics. Theoretical
attempts to explain their origin include primordial
magnetic fields \citep{Bernui:2008ve}, non-trivial cosmic topologies
\citep{Luminet:2003dx,Riazuelo:2003ud}, globally anisotropic models
of the early universe \citep{Pereira:2007yy,Pitrou:2008gk,Gumrukcuoglu:2007bx,Ackerman:2007nb}
as well as local manifestations of cosmic anisotropy \citep{Gordon:2005ai},
and even anisotropic models of dark-energy \citep{Battye:2006mb,Koivisto:2007bp}.
Of course, there is a good chance that these anomalies are
due to astrophysical effects \citep{Abramo:2006hs} or even some residual
instrumental cross-contamination, in which case our universe can still
be easily accommodated in the standard scenario. It is 
a question of utmost concern to decide whether these known anomalies 
(as well as others which may be found in the future)
are isolated statistical flukes, or if they are
due to new physical/astrophysical effects.

Despite its importance and the efforts spent on it, 
we still have no compelling explanation
for the nature of the low-$\ell$ anomalies. The main difficulty is
twofold: first, we still do not know how to optimally separate the
question of gaussianity from that of statistical isotropy (see however
\citep{Land:2005dq} for a first step in this direction.) It is thus
possible that our universe is Gaussian but statistically anisotropic,
statistically isotropic but non-gaussian or even non-gaussian {\em and}
anisotropic. Second, if the universe is neither Gaussian nor statistically
isotropic, then it can be -- from a statistical point of view -- virtually
anything: there is only one kind of gaussianity and isotropy, but
there are infinite ways to brake either one. The absence of
theoretical guidelines will inevitably lead to an infinite number
of models and no underlying symmetries, which still would mean that
we could not account for confirmed anomalies.

This means we must analyze the problem in as much a
model-independent manner as possible. We can, for example, start from
the very basic definition of our statistical quantities (such as
the two-point correlation function) and check
whether they can be modified in a model-independent manner, basing
our reasoning solely on physical symmetries and observational hints.

One such possibility is to consider the two-point temperature correlation
function, $C(\hat{\boldsymbol{n}}_{1},\hat{\boldsymbol{n}}_{2})$, without
some of the symmetries of the underlying space-time. Attempts in this
direction have been made by Pullen and Kamionkowski \citep{Pullen:2007tu},
where the temperature correlation function is assumed to depend on
the direction of any given unit vector in the celestial sphere, in
such a way that one can search for power multipole moments in temperature
maps. However, that approach consists of considering the temperature
correlation function at zero lag, and thus it does not allow us to
consider correlations between two different points in the sky.

Another possibility is to consider the correlation function in its full form,
i.e, a function that depends on all pairs of independent unit vectors
in the sphere $S^{2}$. This idea, which was introduced by Hajian
and Souradeep \citep{Hajian:2003qq,Hajian:2004zn,Hajian:2005jh},
consists in expanding the temperature correlation function in a {\em bipolar}
spherical harmonic series in order to take into account its functional
dependence. The authors then construct a bipolar power spectrum
$\kappa_{\ell}$ which can account for deviations from statistical
anisotropy if observations give us $\kappa_{\ell}>0$ at a statistically
significant level. Unfortunately, that approach is too generic: 
it is not clear what the associated statistical test is measuring,
or how one can motivate it in the absence of an underlying 
theoretical or phenomenological model.

In this work we also go back to the two-point correlation function 
$C(\hat{\boldsymbol{n}}_{1},\hat{\boldsymbol{n}}_{2})$,
but instead ask whether it can depend not only on the separation angle between
two given unit vectors, $\cos\vartheta=\hat{\boldsymbol{n}}_{1}\cdot\hat{\boldsymbol{n}}_{2}$,
but also on the orientation of the plane of the great circle defined by the unit vectors.
Such a functional dependence can be unambiguously constructed once
we realize that for any two unit vectors in the CMB sky, their angular
separation and their associated plane are uniquely defined by their dot and cross
products, respectively. The new planar dependence (on the direction defined by
the normal to the planes of the two unit vectors) codifies modulations of the usual
two-point correlation function as we rotate these planes while keeping the separation
angles $\vartheta$ fixed.

We have also constructed, in a completely model-independent way, an
angular-planar power spectrum and its associated unbiased estimator, 
which naturally generalizes the usual angular power spectrum $C_\ell$, and for which we
recover the known results in the limit of statistical isotropy.\\


Our approach has a strong observational motivation, which lies in the
fact that some astrophysical planes, like the galactic and ecliptic
ones, play an important role in CMB measurements and could still
be manifested in the data if the foregrounds were improperly removed.
One such example was possibly found in \citep{Copi:2005ff} where,
besides the alignment of the multipoles $\ell=2$ and $\ell=3$, the
authors detected a strong correlation between these two and the ecliptic
plane. The existence of a preferential plane could also be related
to the so-called north-south asymmetry \citep{Eriksen:2003db,Eriksen:2007pc,Bernui:2008cr},
in which case a plane could naturally separate regions of maximum
and minimum temperature power. There exists also a third situation
in which a physical plane can play and important role in cosmology,
namely, the unavoidable presence of our galactic plane in all CMB
measurements acts as an important source of astrophysical and foreground
contamination. All these facts lead us to believe that a planar signature
on the correlation function would be an important statistical property of
the CMB, and is a potential test of its nature.\\

We have organized this work in the following way:
we begin $\S$\ref{tcf} with a brief description of the two-point
correlation function and its general properties. After discussing some
of its known generalizations, we extend our argument to include a
planar dependence. In $\S$\ref{angular-planar}
we carry a multipolar decomposition of the correlation function with
planar dependence and show how the resulting coefficients (i.e., the angular-planar 
power spectrum) are related to the usual temperature multipolar coefficients $a_{\ell m}$'s.
This leads us to the question of how to build an unbiased estimator
to measure planar signatures in temperature maps and, in particular,
how this can be implemented with the help of a simple chi-square analysis.
We illustrate, still in this section, the application of our statistics to the well-known 
$\Lambda\mbox{CDM}$ concordance model, where we present some figures for the ``best-fit'' $\Lambda\mbox{CDM}$ angular-planar power spectrum. In $\S$\ref{estimador-aniso} we use
a chi-square test to search for planar signatures in the WMAP full-sky temperature 
maps, and show that the angular scales $\ell=7$ and $\ell=8$ seem to be slightly anomalous 
for a particular range of planar separation $l$. We conclude in $\S$\ref{conclusions}, 
where we also give some perspective of further developments.

\section{Temperature correlation function\label{tcf}}

The main observable in the CMB is the temperature fluctuation field,
$\Delta T $. In its full generality, this field is a function of a
position vector $\hat{\boldsymbol{n}}$ and of the time interval in
which we measure this temperature -- but in practice our measurements 
are made in time intervals which are negligible compared with the 
cosmological timescales. 
The field $\Delta T$ is a scalar, continuous function on the unit sphere, which
means we can decompose it in the usual fashion, in terms of spherical harmonics:
\begin{equation}
\Delta T(\hat{\boldsymbol{n}})=\sum_{\ell,m}a_{\ell m}Y_{\ell m}(\hat{\boldsymbol{n}})\,.
\label{DeltaT}
\end{equation}
All information is therefore encrypted in the multipolar coefficients
$a_{\ell m}$. Essentially all inflationary models predict these coefficients
not as uniquely given, but rather as realizations of a random variable,
in such a way that the physics is not in the $a_{\ell m}$'s themselves,
but rather on their statistical properties. Since, by construction, this field
has zero expectation value, $\langle\Delta T\rangle=0$, the two-point
correlation function expresses the first nontrivial momenta of the
underlying statistical properties of the physical field, and is given by:  
\begin{equation}
C(\hat{\boldsymbol{n}}_{1},\hat{\boldsymbol{n}}_{2})\equiv\langle\Delta T(\hat{\boldsymbol{n}}_{1})
\Delta T(\hat{\boldsymbol{n}}_{2})\rangle=\sum_{\ell_{1},m_{1}}\sum_{\ell_{2},m_{2}}
\langle a_{\ell_{1}m_{1}}a_{\ell_{2}m_{2}}^{*}\rangle Y_{\ell_{1}m_{1}}(\hat{\boldsymbol{n}}_{1})
Y_{\ell_{2}m_{2}}^{*}(\hat{\boldsymbol{n}}_{2})\,.
\label{full-cf}
\end{equation}
 Alternatively, the \textit{covariance matrix} above, $\langle a_{\ell_{1}m_{1}}a_{\ell_{2}m_{2}}^{*}\rangle$,
gives all the information about the quadratic momenta of the underlying
distribution. If the field $\Delta T$ is Gaussian, then this covariance
matrix encloses {\em all} the information that is needed to describe the
nature of the fluctuation field (\ref{DeltaT}). In this
work we shall restrict ourselves to a fiducial Gaussian model, for simplicity.

We note also that the separable nature of the definition (\ref{full-cf})
implies a reciprocity relation for the correlation function: 
\begin{equation}
C(\hat{\boldsymbol{n}}_{1},\hat{\boldsymbol{n}}_{2})=C(\hat{\boldsymbol{n}}_{2},\hat{\boldsymbol{n}}_{1})\,.
\label{rr}
\end{equation}
This symmetry must always be satisfied, regardless of the underlying
physics.

\subsection{Isotropic case\label{Isotropic-case}}

In a globally homogeneous and isotropic universe, the two-point correlation
function of the temperature can only depend on the separation angle
between the vectors $\hat{\boldsymbol{n}}_{1}$ and $\hat{\boldsymbol{n}}_{2}$,
that is: 
\begin{eqnarray}
C(\hat{\boldsymbol{n}}_{1},\hat{\boldsymbol{n}}_{2}) 
& = & C(\hat{\boldsymbol{n}}_{1}\cdot\hat{\boldsymbol{n}}_{2})\nonumber \\
& = & \sum_{\ell}\frac{2\ell+1}{4\pi}\, C_{\ell}\, P_{\ell}(\hat{\boldsymbol{n}}_{1}\cdot\hat{\boldsymbol{n}}_{2})\,.
\label{fc-iso}
\end{eqnarray}
 Comparing this expression with Eq. (\ref{full-cf}), we notice that
the covariance matrix becomes diagonal: 
\begin{equation}
\langle a_{\ell_{1}m_{1}}a_{\ell_{2}m_{2}}^{*}\rangle=C_{\ell_{1}}\delta_{\ell_{1}\ell_{2}}\delta_{m_{1}m_{2}}\,,
\label{cov-matrix}
\end{equation}
 with the diagonal terms given by the \textit{angular power spectrum},
$C_{\ell}$. In principle the angular power spectrum suffices to describe
the statistical properties of the temperature field (\ref{DeltaT}).
However, since we have only one universe to measure, and therefore
only one set of $a_{\ell m}$'s, the average in (\ref{cov-matrix})
is poorly determined. The best we can do then is to take advantage
of the ergodic hypothesis, which states that averaging over an ensemble
can be treated as averaging over space, and hence to consider each
of the $2\ell+1$ real numbers in $a_{\ell m}$ as statistically independent,
in such a way as to build a statistical estimator for the $C_{\ell}$'s:
\[\widehat{C}_{\ell}\equiv\frac{1}{2\ell+1}\sum_{m=-\ell}^{\ell}|a_{\ell m}|^{2}\,.\]

Since $\langle\widehat{C}_{\ell}\rangle=C_{\ell}$, this estimator
is said to be unbiased. Also, because for a Gaussian field
$\langle(\widehat{C}_{\ell}-C_{\ell})(\widehat{C}_{\ell'}-C_{\ell'})\rangle\propto\delta_{\ell\ell'}$,
this estimator has the least {}``cosmic variance''. $\widehat{C}_{\ell}$ is, therefore,
the best estimator that can measure the statistical properties
of the multipolar coefficients $a_{\ell m}$ when both statistical
isotropy and gaussianity hold.

\subsection{Some anisotropic cases\label{anisotropic-case}}

The first line in Eq. (\ref{fc-iso}) for the temperature two-point correlation function
is valid if and only if the universe is statistically isotropic. This
means that any functional dependence that does not reduce to a dependence
on $\cos\vartheta=\hat{\boldsymbol{n}}_{1}\cdot\hat{\boldsymbol{n}}_{2}$
will measure some deviation from statistical isotropy. There are infinite
possible combinations of $\hat{\boldsymbol{n}}_{1}$ and $\hat{\boldsymbol{n}}_{2}$
that violate statistical isotropy. However, since the vectors $\hat{\boldsymbol{n}}_{1}$
and $\hat{\boldsymbol{n}}_{2}$ are constrained to have a common origin
and size, symmetry and simplicity does not leave us many choices.
One possibility is to consider these two vectors as being the same,
in which case we are left with a correlation function of the form:
\begin{equation}
C:S^{2}\rightarrow\mathbb{R}\,,\label{S2R}
\end{equation}
 and for which a decomposition similar to (\ref{DeltaT}) exists.
This form of the correlation function makes it suitable for searching
for power multipole moments in CMB temperature and polarization maps,
once we define a power multipole moment estimator \citep{Pullen:2007tu}.
On the other hand, this is also a correlation function at zero lag,
so by construction it does not allow us to consider anisotropic correlations
between different points in the sky.

A second possibility is to consider the correlation function as being
the most general (but separable) function of two unit vectors that
one can possibly have \citep{Hajian:2003qq}: 
\begin{equation}
C:S^{2}\times S^{2}\rightarrow\mathbb{R}\,.\label{S2S2R}
\end{equation}
 This function admits a decomposition in terms of the bipolar spherical
harmonics \cite{Hajian:2003qq} which has the nice property of behaving
-- in many mathematical aspects -- as the usual spherical harmonics.
The main drawback of the decomposition (\ref{S2S2R}), however, is
that it carries too many degrees of freedom which, in the absence
of a specific cosmological model, cannot be resolved with simple estimators.
Therefore, these two approaches are either too simple or too generic
to reveal deviations from statistical isotropy in a more model-independent
way.

\subsection{Anisotropy through planar dependence}

The guiding principle used in the construction of (\ref{S2R}) and
(\ref{S2S2R}) is rather general and based mainly on our prejudices
about what statistical anisotropy should look like. However, in the
absence of theoretical guidelines, we have to confine ourselves to
the observations of the CMB temperature or, more specifically, to
the signature of its known anomalies. One example is the role played
by the galactic and ecliptic plane in the quadrupole-octupole/north-south
anomalies \citep{deOliveiraCosta:2003pu,Eriksen:2003db,Schwarz:2004gk,Land:2005ad,Eriksen:2007pc},
not to mention the importance of our galactic plane as a source of
foreground contamination in the construction of cleaned CMB maps.
The existence of a cosmic plane might even be a manifestation of some
mirror symmetry \citep{Land:2005jq}.

\begin{figure}[H]
\begin{centering}
\includegraphics[clip,scale=0.4]{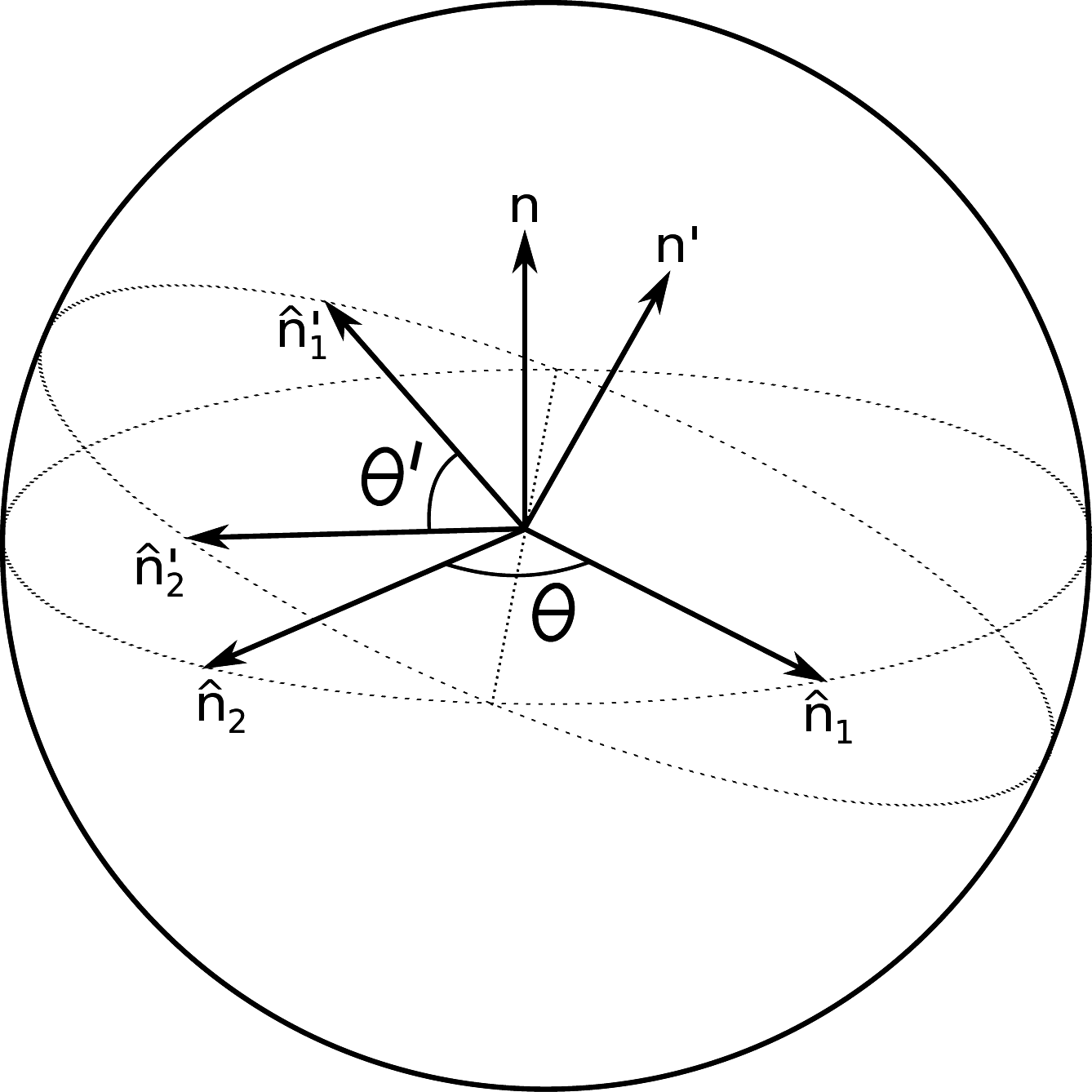} 
\par\end{centering}

\caption{Schematic representation of the functional dependence of the correlation
function (\ref{fc-aniso-formal}). In the anisotropic case we are
considering, fields at the positions given by vectors living on different
planes can have different correlations, regardless of their angular
separation.\label{esfera}}

\end{figure}

In general, the simple fact that we are bound to make all our measurements
inside our galactic plane suggests that the correlation between fields
at two positions $\hat{\boldsymbol{n}}_{1}$ and $\hat{\boldsymbol{n}}_{2}$
might be sensitive not only to their separation angle but also to
the orientation of the plane they live in, as is shown in Fig. \ref{esfera}.

Such a planar dependence can be included in the correlation function
if we realize that two unit vectors on the sphere $S^2$ uniquely define 
both a separation angle $\vartheta$ and a direction $\hat{\mathbf{n}}$
perpendicular to the great circle (or plane) where they live. 
We are then left with a new possibility for the functional
dependence of the two-point correlation function: 
\begin{equation}
C(\hat{\boldsymbol{n}}_{1},\hat{\boldsymbol{n}}_{2})=C(\hat{\boldsymbol{n}}_{1}\times\hat{\boldsymbol{n}}_{2})\,,
\label{fc-aniso-formal}
\end{equation}
which corresponds formally to a function of the form $C:D^{3}\rightarrow\mathbb{R}$,
where $D^{3}$ is the set of all points $(x,y,z)$ such that $x^{2}+y^{2}+z^{2}\leq1$
\footnote{This is the three-dimensional version of the familiar two-dimensional disc, 
for which the quotient with the unit circle gives the 2-sphere: $S^2=D^2/S^1$.}.
Defining $\mathbf{n}\equiv\hat{\boldsymbol{n}}_{1}\times\hat{\boldsymbol{n}}_{2}$,
the above expression can be further decomposed in spherical coordinates
as follows: 
\begin{equation}
C(\mathbf{n})=\sum_{\ell}\sum_{l,m}\frac{2\ell+1}{\sqrt{4\pi}}\mathcal{C}_{\ell}^{lm}
P_{\ell}(\cos\vartheta)Y_{lm}(\hat{\mathbf{n}})\,,\quad l\in2\mathbb{N}\,,
\label{fc-aniso}
\end{equation}
where: 
\[|\mathbf{n}|=\sin\vartheta\,,\quad\hat{\mathbf{n}}=\{\theta,\phi\}\,.\]
 Notice that: 
\begin{equation}
\hat{\boldsymbol{n}}_{1}\cdot\hat{\boldsymbol{n}}_{2}=\cos\vartheta=\cos\theta_{1}\cos\theta_{2}
+\sin\theta_{1}\sin\theta_{2}\cos(\phi_{1}-\phi_{2})\,,
\label{cos-theta}
\end{equation}
 where $(\theta_{i},\phi_{i})$ are the angles defined by the vectors
$\hat{\boldsymbol{n}}_{i}$.

Some comments on the decomposition (\ref{fc-aniso}) are in order.
First, we note that there is an intrinsic ambiguity in the sense of
the vector $\mathbf{n}$ (as we might as well have defined 
$\mathbf{n}\equiv\hat{\boldsymbol{n}}_{2}\times\hat{\boldsymbol{n}}_{1}$),
which is obviously inherited from the ambiguity in the definition
of the normal to a plane. This ambiguity can be avoided if we restrict
the sum in $l$ to even values, which is what we will do from now
on. Note that such a restriction arises naturally as a consequence
of the reciprocity relation Eq. (\ref{rr}). Second, for $\ell=0$
we recover (\ref{S2R}) and therefore all the analysis made in \citep{Pullen:2007tu}
arises as a special case here.

\section{Angular-planar power spectrum\label{angular-planar}}

The multipolar $\mathcal{C}_{\ell}^{lm}$ coefficients in Eq. (\ref{fc-aniso})
correspond to a generalization of the usual angular power spectrum
$C_{\ell}$'s. In fact, they can be seen as a spherical harmonic decomposition
of the angular power spectrum, if it suffers modulations as we sweep planes
on the sphere.
The function $C_{\ell}(\hat{\mathbf{n}})$ for a given $\ell$ is:
\[C_{\ell}(\hat{\mathbf{n}})=\sqrt{4\pi}\sum_{l,m}\mathcal{C}_{\ell}^{lm}Y_{lm}(\hat{\mathbf{n}})\,,
\quad l\in2\mathbb{N}\,.\]
 Clearly, the monopole of $C_{\ell}(\hat{\mathbf{n}})$ (the average
over the whole sphere) is the usual angular power spectrum, $\mathcal{C}_{\ell}^{00}=C_{\ell}$,
and the higher multipoles measure modulations of the spectrum.

Since we are restricting our analysis
to the Gaussian case, the set of coefficients $\mathcal{C}_{\ell}^{lm}$
completely characterizes the two-point correlation function. Still,
what is accessible through observations are temperature maps which
we can use to try to estimate the correlation function. In this respect
the multipolar coefficients $\mathcal{C}_{\ell}^{lm}$ would be of
limited interest, unless we can relate them directly to our observables.
It would be interesting if we could, for example, relate these coefficients
to the covariance matrix $\langle a_{\ell_{1}m_{1}}a_{\ell_{2}m_{2}}^{*}\rangle$
by equating expressions (\ref{fc-aniso}) and (\ref{full-cf}), as
is usually done. However, this procedure is far from being trivial,
since the complicated coupling of the angles $\vartheta$, $\theta$
and $\phi$ defined in (\ref{cos-theta}) make it difficult to use
the usual orthogonality relations to isolate the $\mathcal{C}_{\ell}^{lm}$'s.

Fortunately, as we show below, we can estimate the $\mathcal{C}_{\ell}^{lm}$'s
if we use the invariance of the scalar product $\hat{\boldsymbol{n}}_{1}\cdot\hat{\boldsymbol{n}}_{2}$
and chose our coordinate system in order to integrate out the $\vartheta$
dependence. Once this is done, we make a passive rotation of the coordinate
system and then we integrate over the remaining angles $\theta$ and
$\phi$, which then are given precisely by the Euler angles used in
the rotation. The details are rather technical and can be found in
the Appendix. The final expression is: 
\begin{equation}
\frac{\mathcal{C}_{\ell}^{lm}}{\sqrt{2l+1}}=2\pi\sum_{\ell_{1},m_{1}}\sum_{\ell_{2},m_{2}}
\langle a_{\ell_{1}m_{1}}a_{\ell_{2}m_{2}}\rangle\left(\begin{array}{ccc}
l & \ell_{1} & \ell_{2}\\
m & m_{1} & m_{2}\end{array}\right)I_{\ell_{1}\ell_{2}}^{l,\ell} \, ,
\label{Clmell}
\end{equation}
where the 6-index expression in parenthesis is the Wigner 3J symbol, and
 \begin{equation}
I_{\ell_{1}\ell_{2}}^{l,\ell}\equiv\sum_{m}(-1)^{m}\lambda_{\ell_{1}m}\lambda_{\ell_{2}m}
\left(\begin{array}{ccc}l & \ell_{1} & \ell_{2}\\
0 & m & -m\end{array}\right)\int_{0}^{\pi} d(-\cos{\vartheta})\, 
P_{\ell}(\cos\vartheta)e^{im\vartheta}\,,
\label{Int-l-ell}
\end{equation}
where $\lambda_{\ell_{i}m}$ are a set of coefficients resulting
from the $\vartheta$ integration, which vanish unless $\ell_{i}+m=$
even (see the Appendix for more details.)

It is easy to show that expression (\ref{Clmell}) induces no coupling between
the eigenvalues $\ell$ and $l$, as expected, since the length of
the vector $\mathbf{n}$ is completely independent of its orientation.
There are, however, subtle couplings present in (\ref{Clmell}) which
do make a difference when we apply it to real data. This is due to
the Legendre polynomial in the integral (\ref{Int-l-ell}), which
selects only those values of $\ell_{1}$ and $\ell_{2}$ which have
the same parity as the angular momentum $\ell$. Moreover, the 3J
symbols appearing in (\ref{Clmell}) give different weights to the
triple $(l,\ell_{1},\ell_{2})$ depending on the parity of $(\ell_{1},\ell_{2})$
and, as a consequence, we can expect typical oscillations in any function
of (\ref{Clmell}) that we may build when plotted as a function of
$\ell$. This will be shown explicitly in the next section, when we
apply these tools to the WMAP 5 year data.\\

Expression (\ref{Clmell}) does not take into account the fact that
real data is not given exactly by (\ref{DeltaT}), but rather by a
pixelized temperature map which is a combination of the true cosmological
signal, plus instrumental noise and residual foreground contamination.
Schematically, the temperature of the map in each pixel $i$ is given
by 
\[\Delta T^{\,\scriptsize{\mbox{map}}}(\hat{\boldsymbol{n}}_{i})
=\Delta T^{S}(\hat{\boldsymbol{n}}_{i})+\Delta T^{N}(\hat{\boldsymbol{n}}_{i})+\Delta T^{R}(\hat{\boldsymbol{n}}_{i})\,.\]
 Typically, the cosmological signal $\Delta T^{S}$ is smoothed out
by a Gaussian beam $W(\hat{\boldsymbol{n}}_{i})$ of finite width
which, in harmonic space, is given by $W_{\ell}=\exp(-\ell^{2}\sigma_{b}^{2})$,
where $\sigma_{b}=\theta_{\scriptsize{\mbox{fwhm}}}/\sqrt{8\ln2}$
and $\theta_{\scriptsize{\mbox{fwhm}}}$ is the beam full width at
half maximum. For the V-band frequency map of the WMAP experiment,
$\theta_{\scriptsize{\mbox{fwhm}}}=0.35^{\circ}$, which implies a
minimum $\ell_{\min}\gtrsim390$ for which the effect of a beam smoothing
will be important, much higher than the low-$\ell$ regions where
known anomalies were reported. Thus, for the sake of simplicity we
will neglect the effect of the beam in this work. Also, for the $\ell\lesssim390$
region, cosmic variance is known to dominate the source of error over
instrumental noise, and therefore we can neglect the latter as well.
On the other hand, the residual foreground can be an important source
of contamination, and therefore deserves a careful analysis which
is beyond the scope of the present work. In a companion paper we carry
a more rigorous analysis of planar signature in CMB data in which
the effect of the residual foreground will be estimated \citep{NewEntry1}.

\subsection{Statistical estimators and $\chi^2$ analysis}

We now would like to use expression (\ref{Clmell}) to examine the
observed universe. We start by noting that in the limit of statistical
isotropy (SI), that is, when 
$\langle a_{\ell_{1}m_{1}}a_{\ell_{2}m_{2}}\rangle=C_{\ell_{1}}\delta_{\ell_{1}\ell_{2}}\delta_{m_{1}m_{2}}$,
expression (\ref{Clmell}) reduces to 
\begin{equation}
\mathcal{C}_{\ell}^{lm}\;\overset{\textrm{(SI)}}{=}\; C_{\ell}\delta_{l0}\delta_{m0}\,.
\label{SI}
\end{equation}
 Conversely, if the only non-zero $\mathcal{C}_{\ell}^{lm}$'s are
given by $l=m=0$, then $\mathcal{C}_{\ell}^{00}=C_{\ell}$. Therefore,
statistical isotropy is achieved if and only if the $\mathcal{C}_{\ell}^{lm}$'s
are of the form (\ref{SI}), and any observational deviation from
this relation would be an indication of statistical anisotropy.

However, we only get to observe one universe, and this makes the ``cosmic sample variance'' 
a severe restriction that we have to live with. This means that if we want to know, let's
say, the mean value and variance of the $\mathcal{C}_{\ell}^{lm}$'s,
we will have to build statistical functions which can only 
{\em estimate} these properties, just like it is done with the fundamental
quantities $a_{\ell m}$ and the associated estimators $\widehat{C}_{\ell}$
(see the discussion in $\S$\ref{Isotropic-case}). 

In other words, in order to evaluate the statistical properties of
the $\mathcal{C}_{\ell}^{lm}$'s we will have to treat them as our
new ``fundamental'' quantities, which will be determined exclusively
as a function of the $a_{\ell m}$'s. As a consequence, we will redefine
expression (\ref{Clmell}) as: 
\begin{equation}
\mathcal{C}^{lm}_\ell\;\rightarrow\;2\pi\sqrt{2l+1}\sum_{\ell_{1},m_{1}}
\sum_{\ell_{2},m_{2}}a_{\ell_{1}m_{1}}
a_{\ell_{2}m_{2}}\left(\begin{array}{ccc}l & \ell_{1} & \ell_{2}\\
m & m_{1} & m_{2}\end{array}\right)I_{\ell_{1}\ell_{2}}^{l,\ell} \; ,
\label{Clmell-hat}
\end{equation}
and will treat the coefficients $\mathcal{C}_{\ell}^{lm}$ as uniquely
given once we have a map. Of course, expression (\ref{Clmell-hat})
is nothing more than the unbiased estimator of the angular-planar 
power spectrum (\ref{Clmell}) and -- as long as cosmic variance is an issue -- this 
``second order'' approach we are adopting here (i.e, the prescription of adopting this
estimator of the correlation function as our fundamental quantity,
rather than the temperature field) is the best we can do when searching
for statistical deviations of isotropy. In theory, it is also possible
to use the CMB polarization induced by galactic clusters to probe
different surfaces where CMB photons last scattered, and to use such
independent measurements as a way to alleviate cosmic variance \citep{Kamionkowski:1997na,Abramo:2006gp}.
However, the gain in terms of a reduced variance is still limited.\\

Having these limitations in mind, we can now ask: how good does a theoretical model of anisotropy,
$\mathcal{C}^{{\rm th},lm}_\ell$, fit the observational data once it is given by (\ref{Clmell-hat})? To answer this question we can use the well-known chi-square ($\chi^2$) goodness-of-fit test, which in our case can be written in the following ``generalized'' form:
\[
(\chi^{2})_{\ell}^{l}\equiv\sum_{m=-l}^{l}
\frac{|\mathcal{C}_{\ell}^{lm}-\mathcal{C}_{\ell}^{{\rm th},lm}|^{2}}
{(\sigma_{\ell}^{lm})^{2}} \; ,
\]
in which the scales $l$ and $\ell$ are seem as independent degrees of freedom, and where 
$\sigma^{lm}_\ell$ is just the standard deviation of the difference 
$({\cal C}^{lm}_{\ell}-{\cal C}^{{\rm th},lm}_{\ell})$. Although this expression can be 
readily applied to any theoretical model of anisotropy, in practice it is better to work 
with its reduced version:
\begin{equation}
(\chi_{\nu}^{2})_{\ell}^{l}\equiv\frac{1}{2l+1}\sum_{m=-l}^{l}
\frac{|\mathcal{C}_{\ell}^{lm}-\mathcal{C}_{\ell}^{{\rm th},lm}|^{2}}
{(\sigma_{\ell}^{lm})^{2}} \; ,
\label{chi2}
\end{equation}
which is just the chi-square function divided by the $2l+1$ planar degrees of freedom. \\

Expression (\ref{chi2}) will be the starting point of our statistical analysis, which we will 
pursue in detail in $\S$\ref{estimador-aniso}. Before we move on, it is important to choose a particular cosmological model of anisotropy against which we want to compare our estimator (\ref{Clmell-hat}).


\subsection{$\Lambda$CDM model}

The most important model to be analyzed using the estimator (\ref{Clmell-hat}) is,
of course, the concordance $\Lambda\mbox{CDM}$ model which was confirmed with striking 
accuracy by the 5 year release dataset of the WMAP team \citep{Komatsu:2008hk}.
For this model, statistical isotropy holds and, as we have shown, any multipolar coefficient 
$\mathcal{C}_{\ell}^{lm}$ with non-zero planar dependence should be identically
zero in this case 
[see expression (\ref{SI}).] Therefore, for this model we can take:
\begin{equation}
\mathcal{C}^{{\rm th},lm}_\ell\equiv0\,,
\label{nh}
\end{equation}
where it should be clear that we are only considering the cases with $l\geq2$ 
(as we will do from now on). Now, if the data under analysis is really Gaussian and SI, 
then its covariance matrix can be explicitly calculated:
\begin{eqnarray*}
{\mathcal{M}}^{l l' m m'}_{\ell} & \equiv &
\langle(\mathcal{C}_{\ell}^{lm}-\mathcal{C}_{\ell}^{{\rm th},lm})^{*}
(\mathcal{C}_{\ell}^{l'm'}-\mathcal{C}_{\ell}^{{\rm th},l'm'})
\rangle 
\\
& = & \langle(\mathcal{C}_{\ell}^{lm})^{*}\mathcal{C}_{\ell}^{l'm'}
\rangle = 8\pi^{2}\sum_{\ell_{1},\ell_{2}}C_{\ell_{1}}C_{\ell_{2}}
\left(I_{\ell_{1}\ell_{2}}^{l,\ell}\right)^{2}\delta_{ll'}\delta_{mm'}
\label{cov-matrix-Clmell}
\end{eqnarray*}
where we have used (\ref{cov-matrix}) and the null hypothesis (\ref{nh}).


This covariance matrix has some interesting properties: first, we note that the
planar degrees of freedom in (\ref{chi2}) are really independent in this case. Moreover,
the variance $(\sigma^{lm}_\ell)^2 = {\mathcal{M}}^{l l m m}_\ell$ becomes $m$-independent:
\begin{equation}
\left( \sigma^{lm}_{\ell} \right)^2 
\; \rightarrow \left( \sigma^{l}_\ell \right)^2=
8\pi^{2}\sum_{\ell_{1},\ell_{2}}C_{\ell_{1}}C_{\ell_{2}}
\left(I_{\ell_{1}\ell_{2}}^{l,\ell}\right)^{2}\,.
\label{Blell}
\end{equation}
Second, its diagonal terms (i.e., $\sigma^l_\ell$) are completely determined by the 
angular power spectrum $C_{\ell}$ , 
up to some geometrical coefficients 
which arise as a consequence of the way in which we split
our CMB sky. This makes it possible to give a visual interpretation of the 
angular-planar power spectrum $\mathcal{C}^{lm}_\ell$, similar to that of 
the $C_{\ell}$'s. For that, let us 
introduce the {\em reduced angular-planar spectrum}: 
\begin{equation}
H_{\ell}^{l}\equiv\sqrt{\frac{2\ell+1}{2}} \, \sigma_{\ell}^{l}
\label{Hlell}
\end{equation}
which has a simple interpretation when compared to the usual angular
spectrum, because $H_{\ell}^{0}=\sqrt{(2\ell+1)/2} \sigma_{\ell}^{0} = C_{\ell}$,
as can be easily shown using Eqs. (\ref{Blell}) and (\ref{IL-zero}).

\begin{figure}[h]
\begin{centering}
\includegraphics[scale=0.75]{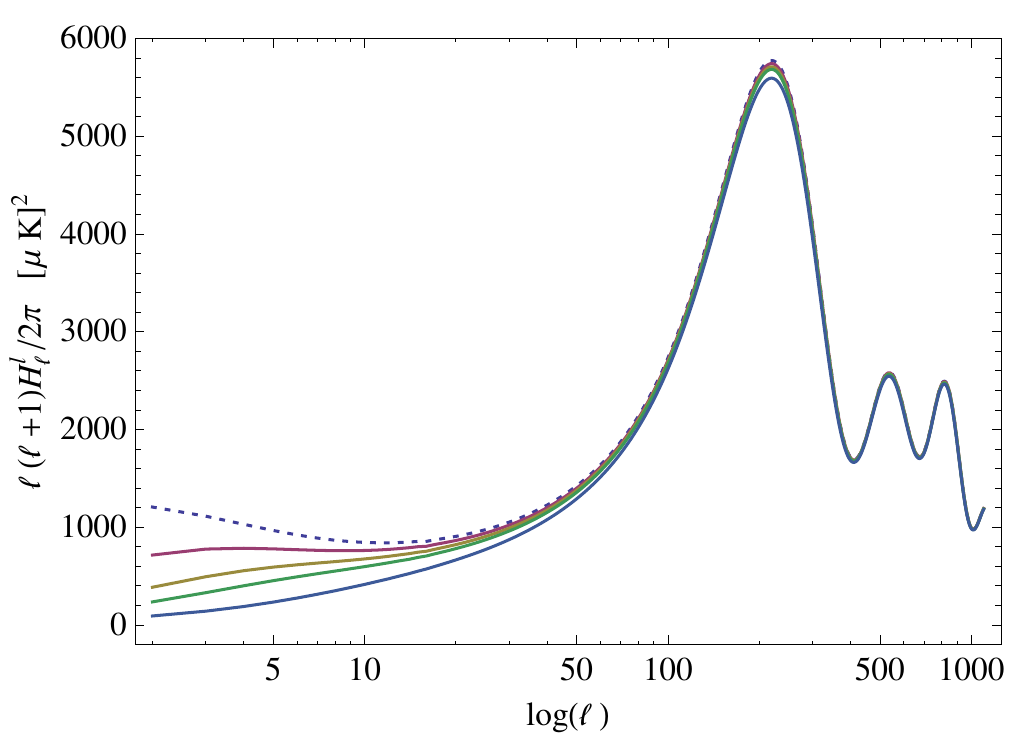} \includegraphics[scale=0.75]{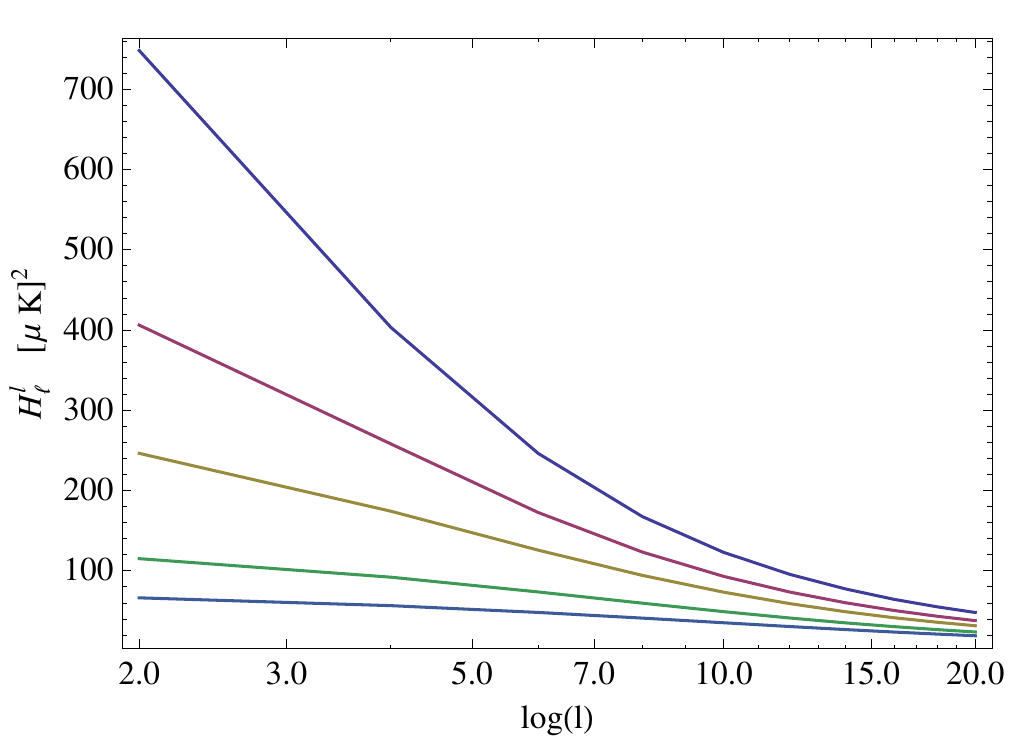} 
\par\end{centering}

\caption{Reduced angular-planar power spectrum $H_{\ell}^{l}$ for a fiducial
$\Lambda\mbox{CDM}$ model. The left panel shows the spectrum as a
function of $\ell$ for the particular values (from top to bottom)
$l=(0,2,4,6,12)$; the case $l=0$ (dotted line) represents the standard
angular power spectrum $C_{\ell}$. The right panel shows the same
spectra, but now as a function of $l$, for the particular values
$\ell=(2,3,4,6,8)$. \label{Hlellfig}}

\end{figure}

In Fig. \ref{Hlellfig} we show some plots of the reduced spectrum
$H_{\ell}^{l}$, both as a function of $l$ and $\ell$. Notice that,
as a result of our planar splitting of the CMB sky, the low-$\ell$ sector
of the spectrum $H_{\ell}^{l}$ is suppressed when we consider planes
separated by smaller angles (bigger values of $l$). This is a consequence
of the nontrivial coupling of the moments $l$, $\ell_{1}$ and $\ell_{2}$:
since the $C_{\ell}$'s are roughly given by a monotonically decreasing
sequence, and since $|l-\ell_{1}|\leq\ell_{2}\leq l+\ell_{1}$, bigger
values of $l$ make the moment $\ell_{2}$ probe deeper and deeper
regions of the Sachs-Wolfe plateau. This suppression reaches cosmological
scales up to the first acoustic peak, after which the planar dependence
becomes negligible.\\

%



\section{$\chi^2$ test of statistical anisotropy\label{estimador-aniso}}

We now come back to the question of how the angular-planar power spectrum 
$\mathcal{C}^{lm}_\ell$ fits the observed universe. We begin by showing that if we 
want to compare our data against the standard $\Lambda\mbox{CDM}$ universe, 
then the chi-square function (\ref{chi2}) becomes a very simple expression. 
As we have shown in the preceding section, for this model, $\sigma^{lm}_\ell=\sigma^l_\ell$ and 
$\mathcal{C}^{{\rm th},lm}_\ell=0$. Therefore (\ref{chi2}) simplifies to:
\begin{equation}
(\chi^2_\nu)^l_\ell=\frac{1}{2l+1}\sum_{m=-l}^l
\frac{|\mathcal{C}^{{\rm obs},lm}_\ell|^2}{(\sigma^l_\ell)^2}\,,
\label{chi2-iso}
\end{equation}
where $\mathcal{C}^{{\rm obs},lm}_\ell$ is calculated by applying the estimator (\ref{Clmell-hat}) to the data given by $a_{\ell m}^{\rm obs}$. It is now clear that 
if the data under analysis is really Gaussian and statistically isotropic, then it 
should be true that:
\[
\langle(\chi^2_\nu)^l_\ell\rangle=\frac{1}{2l+1}\sum_{m=-l}^l
\frac{\langle(\mathcal{C}^{{\rm obs},lm}_\ell)^*\mathcal{C}^{{\rm obs},lm}_\ell\rangle}
{(\sigma^l_\ell)^2}=1\,.
\]
This means that a positive test of planarity will be quantified by how far our chi-square 
function deviates from unity. We can do even better and define a new function as:
\begin{equation}
\overline{\chi}^{\,l}_\ell\equiv(\chi^2_\nu)^l_\ell-1\,
\label{blell-barra}
\end{equation}
which, if significantly different from zero, 
will point towards anisotropy.

It should be stressed that, for a given CMB map, the chi-square analysis 
must be done entirely in terms of that map's data. Indeed, any arbitrary introduction
of a fiducial bias in (\ref{chi2-iso}) (for example, by calculating $\sigma^l_{\ell}$ using 
$C_\ell^{\scriptsize{\Lambda{\rm CDM}}}$) would only include our \textit{a priori} prejudices 
about what the map's anisotropies should look like. The angular spectrum $C_{\ell}$, being 
by construction a measure of statistical isotropy, can only be said to be small/big
when compared to a particular cosmological model (for example, the
$\Lambda\mbox{CDM}$ model). Consequently, an anomalous detection
of $C_{\ell}$ is by no means a measure of statistical anisotropy,
and it is this value that should be used to calculate $\sigma^l_\ell$ if we 
want to find deviations of isotropy, regardless of how high/low it is. Note also that 
while the function $\overline{\chi}^{\,l}_\ell$ has some ``isotropy variance'' 
which could be computed for the $\Lambda$CDM model from first principles, in practice
it is much easier to simulate many realizations of a Gaussian and
isotropic random field to obtain that variance. 

Finally, we would like to mention that although each number
$\overline{\chi}^{\,l}_\ell$ is an individual measure of anisotropy (i.e., planarity),
a consistently biased set of values over a range of $l$'s or $\ell$'s can also 
be seen as an indication of anisotropy, even if all individual $\overline{\chi}_{\ell}^{\,l}$'s in
that range are well within their variance limits.\\

Following the prescription outlined above, we applied the estimator
(\ref{blell-barra}) to the 5 year WMAP full sky data (also known as ILC5 map) where, for
practical reasons, we have restricted our analysis to the range of values $\ell\in[2,12]$ 
and $l\in[2,12]$ (notice that the momenta $l$ can only assume even values). Our results are presented in Fig. \ref{B-barra-ell}, where we keep the momenta $l$ fixed, and vary the 
momenta related to angular separation, $\ell$. As discussed before, for this range
of values cosmic variance dominates over other sources of noise. We 
estimated the effects of cosmic variance by running a simulation of $10^{3}$ realizations 
of this estimator, using the best-fit (theoretical) scalar $C_{\ell}$'s made available in \cite{lambda}; this corresponds to the shaded area in Fig. \ref{B-barra-ell}.

It is also important to explain that, while the data points in Fig. 
\ref{B-barra-ell} were calculated using the ILC5 map {\it alone}, 
we have also included in our analysis a rough 
estimate of the possible residual foreground contamination present in 
the data. This was done 
by computing the sample variances of the full-sky maps shown in Table 
\ref{mapas-ceu-int}, 
which were then used as error bars. 
In other words, the error bars in Fig. \ref{B-barra-ell}
{\it do not} account for instrumental noise, which is
believed to be under control at these angular scales.\\

\begin{center}
\begin{table}[h]
\begin{centering}
\begin{tabular}{cc}
\toprule 
Full sky maps & References\tabularnewline
\midrule
\midrule 
Hinshaw \textit{et. al.} & \cite{Hinshaw:2006ia,Hinshaw:2008kr}\tabularnewline
\midrule
de Oliveira-Costa \textit{et. al.} & \cite{deOliveiraCosta:2006zj}\tabularnewline
\midrule 
Kim \textit{et. al.} & \cite{Kim:2008zh}\tabularnewline
\midrule
Park \textit{et. al.} & \cite{Park:2006dv}\tabularnewline
\midrule
Delabrouille \textit{et. al.} & \cite{Delabrouille:2008qd}\tabularnewline
\bottomrule
\end{tabular}
\par\end{centering}

\caption{Full sky WMAP maps used in our analysis to estimate the error bars
of Fig. \ref{B-barra-ell}. The data points in this (and subsequent) figure 
was calculated using only the ILC5 map \cite{Hinshaw:2006ia,Hinshaw:2008kr}.}
\label{mapas-ceu-int}

\end{table}
\par\end{center}

Theses figures present some peculiarities: first, we notice that the
magnitude of the error bars oscillate for the smallest values of $\ell$.
As we mentioned in $\S$\ref{angular-planar}, this is partially a
consequence of the 3J symbols, which are weights appearing in the
definition of the anisotropic power spectra and whose effect is to
couple differently odd and even multipoles. The second peculiarity
is that, in all these figures, the modulations of the quadrupolar
moment $\ell=2$ are entirely consistent with zero. This result suggests
that the low value of the quadrupole $C_{2}$ is perhaps not a consequence
of statistical anisotropy, at least for the test we are considering
here. Note also that the octupole $\ell=3$, which has been reported
as unusually planar by some groups, grows slightly from $l=4$ to
$l=8$, although it is compatible with cosmic variance in all the
planar range considered.

In what concerns deviations of isotropy, our analysis shows that the
most {}``anomalous'' scales are in the sectors $(l,\ell)=(4,7)$
and $(6,8)$, where we can see that the points $\overline{\chi}_{7}^{4}$
and $\overline{\chi}_{8}^{6}$ are only marginally allowed by the $2\sigma$
cosmic variance area.

\begin{center}
\begin{figure}[H]
\begin{centering}
\includegraphics{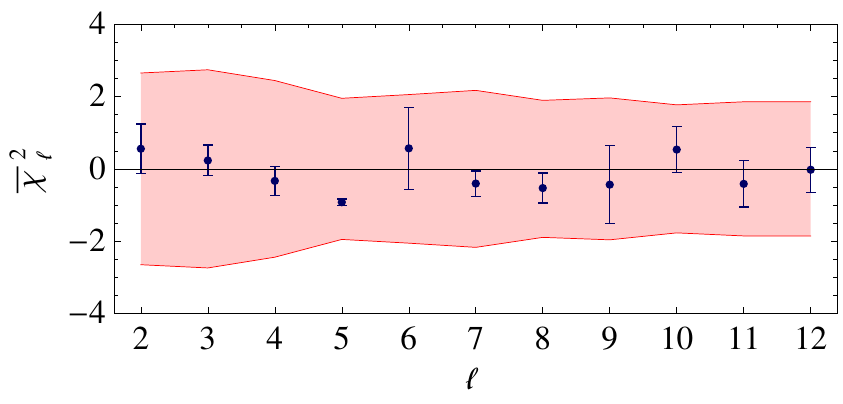} \includegraphics{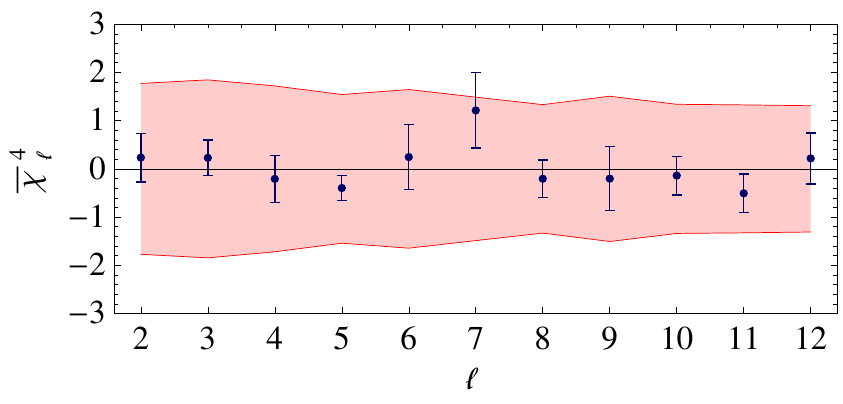} \\
 \includegraphics{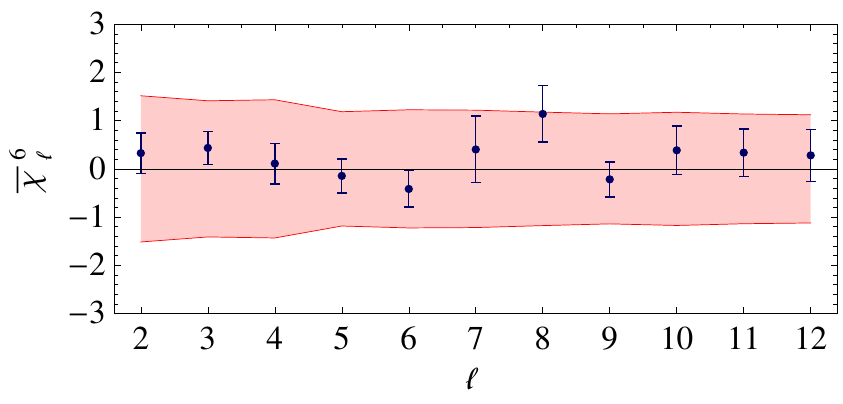} \includegraphics{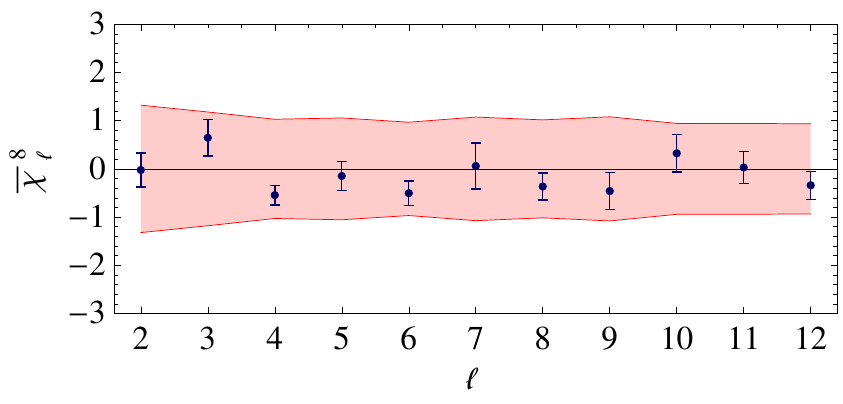} \\
 \includegraphics{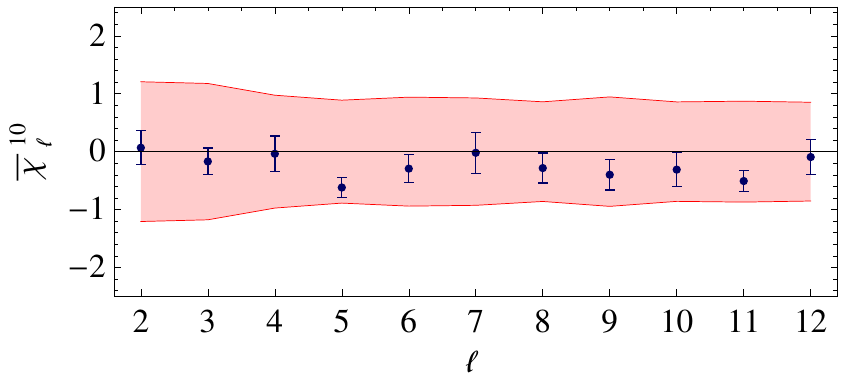} \includegraphics{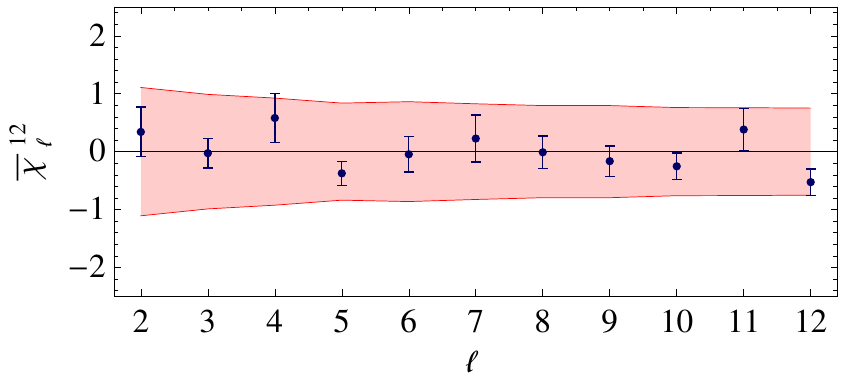} 
\par\end{centering}

\caption{Anisotropic angular-planar estimator applied to the WMAP ILC5 
data. The panel shows $\overline{\chi}_{\ell}^{l}$ as a function of
$\ell$, for the particular values $l=(2,4,6,8,10,12)$. The shaded region 
and error bars represent $2\sigma$ cosmic variance and systematical (foreground) 
contaminations, respectively.
\label{B-barra-ell}}

\end{figure}

\par\end{center}

In order to make the visualization of the above figures easier, we
repeat the analysis but now keeping the angular separation $\ell$
fixed and varying the planar separation $l$. The result is shown
in Fig. \ref{B-barra-l}. Notice that the planar modulations of
the quadrupole $\ell=2$ are consistently positive, but always compatible
with zero. We can also see in these figures the growing behavior in
the octupole $\ell=3$ from $l=4$ to $l=8$ as mentioned before.

\begin{center}
\begin{figure}[H]
\begin{centering}
\includegraphics{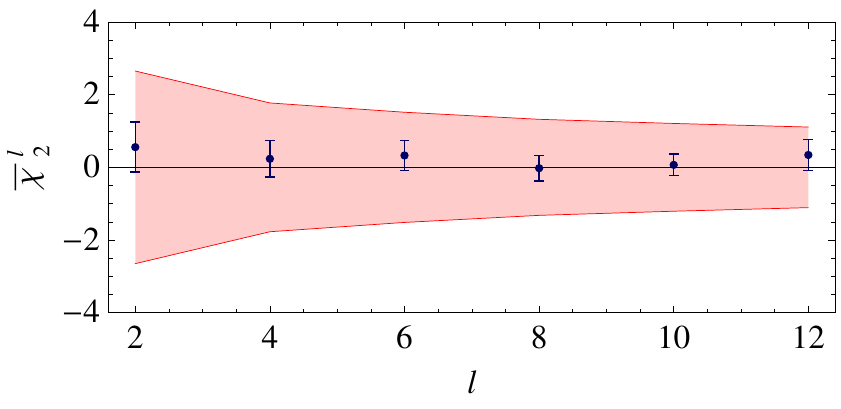} \includegraphics{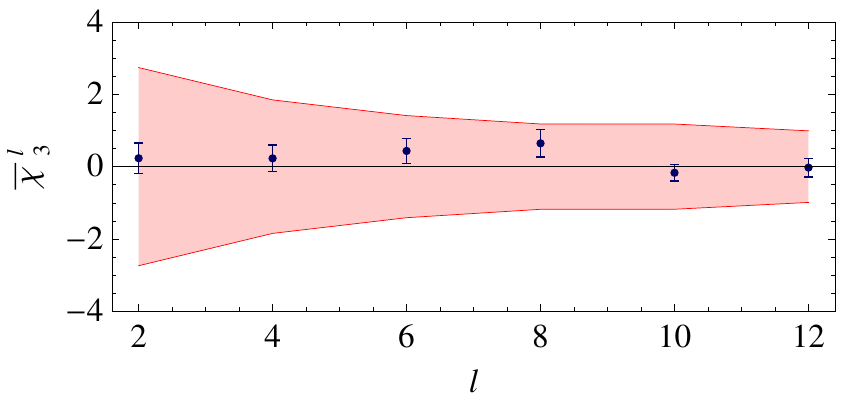} 
\par\end{centering}

\begin{centering}
\includegraphics{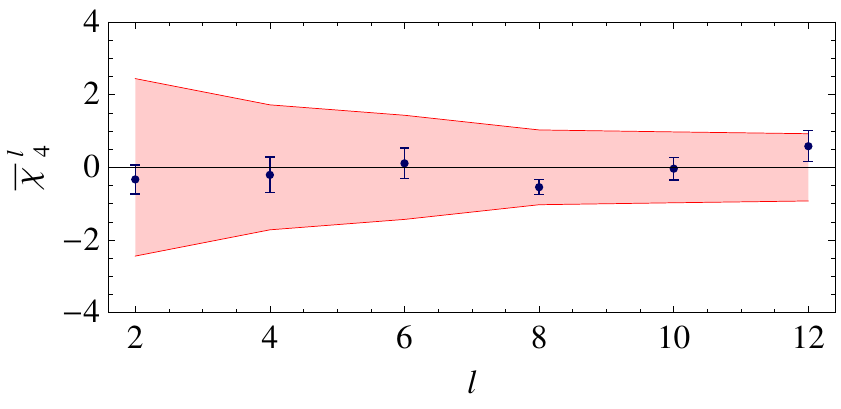} \includegraphics{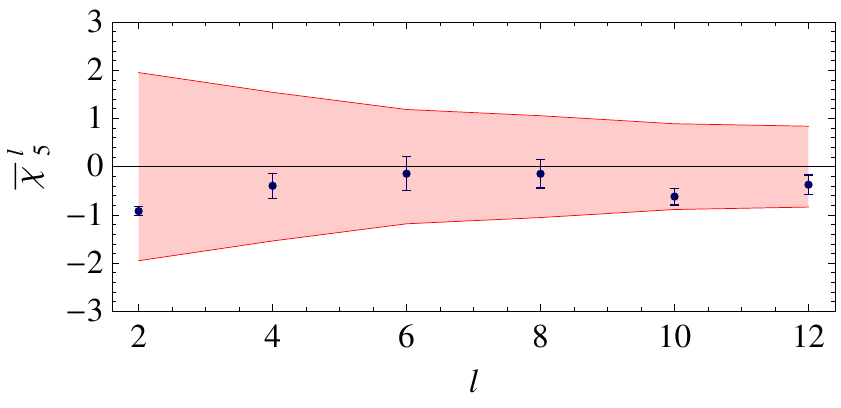}
\par\end{centering}

\begin{centering}
\includegraphics{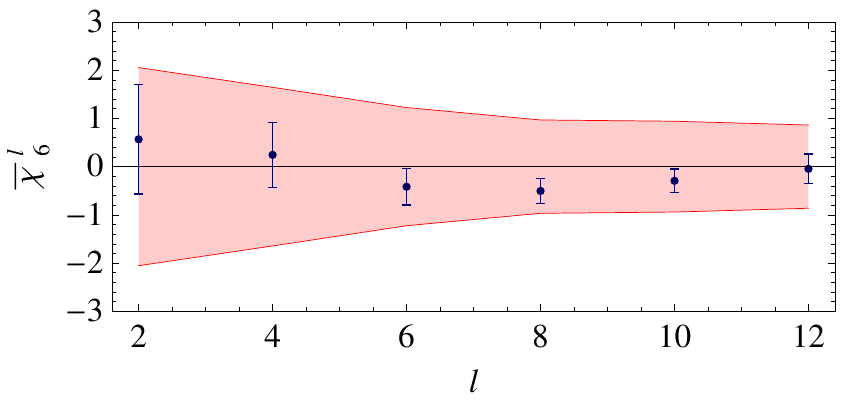} \includegraphics{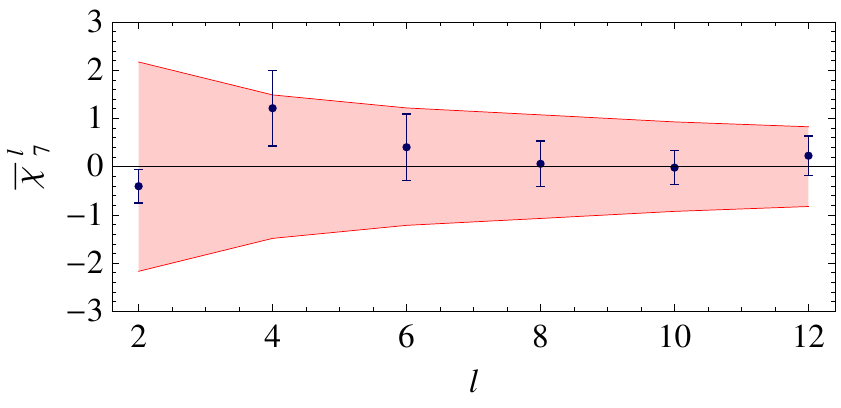}
\par\end{centering}

\begin{centering}
\includegraphics{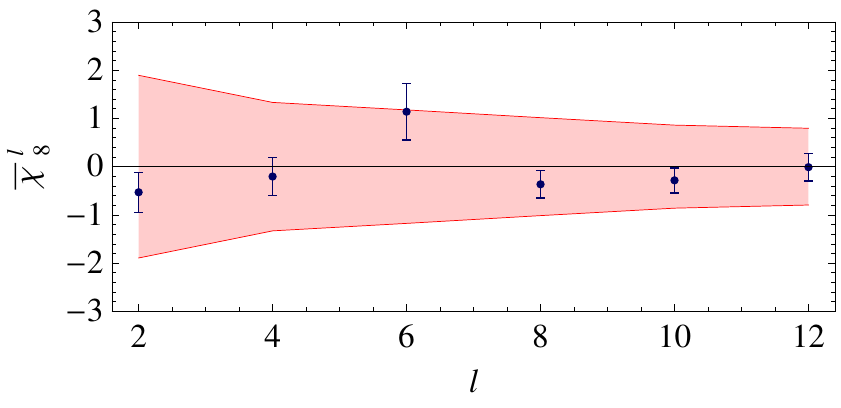} \includegraphics{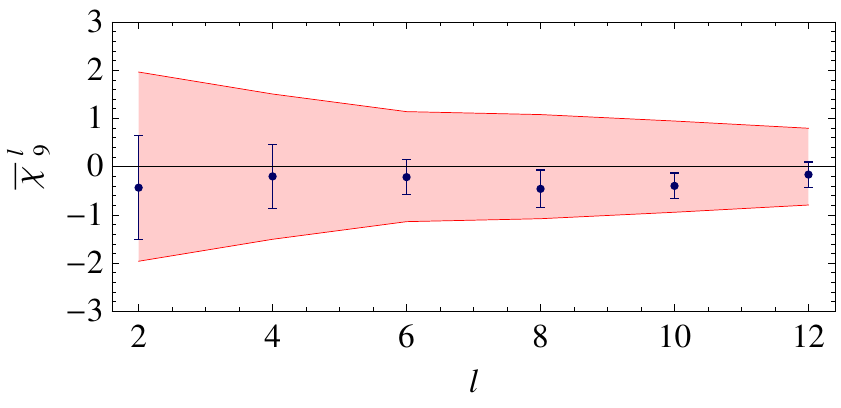}
\par\end{centering}

\begin{centering}
\includegraphics{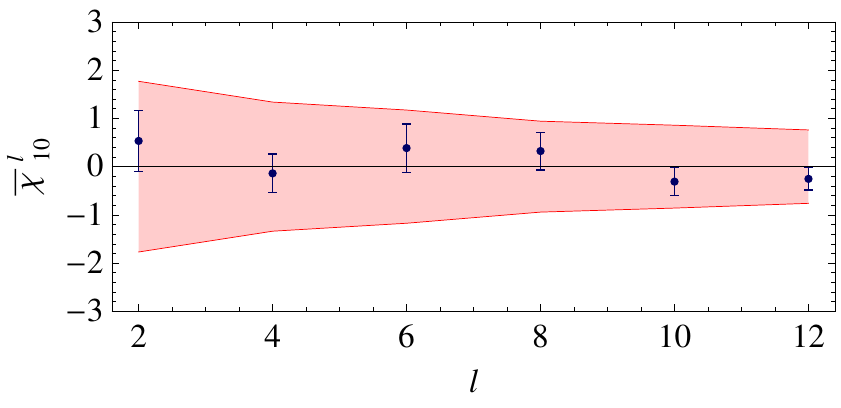} \includegraphics{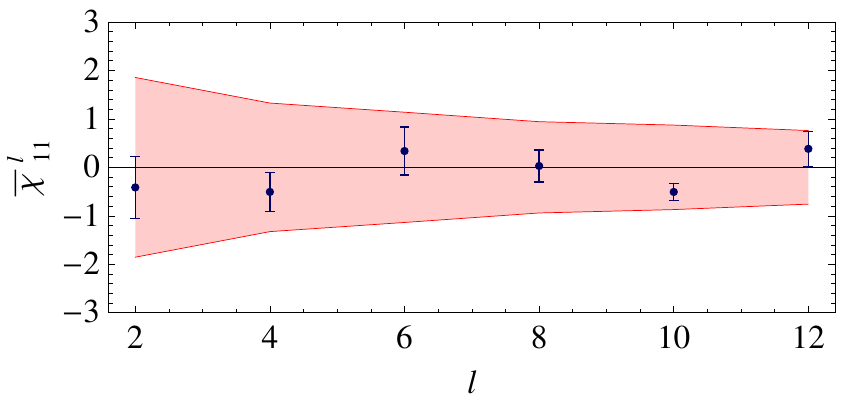}
\par\end{centering}

\caption{Anisotropic angular-planar estimator applied to the WMAP ILC5 
data. The panel shows $\overline{\chi}_{\ell}^{l}$ as a function of
$l$, for the particular values $\ell=(2,3,4,5,6,7,8,9,10,11)$. Note
that for fixed $l$, the error bars do not oscillate (see the text
for more details).\label{B-barra-l}}

\end{figure}

\par\end{center}

\section{Conclusions\label{conclusions}}

We have investigated the minimum statistical framework of modern cosmology by 
enlarging the domain of the two-point correlation
function to admit not only the usual angular dependence, but a directional
(planar) dependence as well. Our observable, the anisotropic angular-planar
power spectrum, can account not only for the usual angular separation
between any two spots in a CMB map, but also for any planar signature 
that this map might have. Besides having a strong observational motivation, an
interesting feature of this approach is that it leads naturally to an unbiased estimator
of statistical anisotropy, in the same spirit as is done with the multipolar temperature
coefficients $a_{\ell m}$'s and their associated estimator $\widehat{C}_{\ell}$.
As an example of its use, we have applied this estimator to a concrete model of cosmology, 
i.e., the $\Lambda\mbox{CDM}$ model, where we have shown that under the hypothesis of 
gaussianity and statistical isotropy, the angular-planar power spectra have zero mean, 
but of course, non-zero covariance.

By means of a simple chi-square analysis, we have also applied our estimator of 
planar anisotropy to the WMAP ILC5 data, where we found that the planar modulations of 
the quadrupole $\ell=2$ are compatible with the null hypothesis 
over the range of planar momenta 
$l\in[2,12]$ we probed. Our results suggests that the low value of the quadrupole $C_{2}$ 
is perhaps due to some local physics, and not to deviations of statistical isotropy,
at least as far as planar modulations are concerned. 

Our analysis has also shown that the angular scales $\ell=7$ and
$\ell=8$ suffer some degree of modulation around the planar scales $l=4$
and $l=6$, respectively. This could be an indication of some foreground
contamination coming from a planar region of typical size $\Delta l=4\sim 6$.
However, a complete treatment of the sources of errors and the effect
of masks is needed before we can reach a more definitive conclusion
-- for that analysis, see \cite{NewEntry1}.

From a theoretical perspective, our techniques can be readily
applied to any particular model of inflation predicting a specific
anisotropic shape for the matter power spectrum. Due to the generality
and simplicity of our formulas, the angular-planar power spectrum can also 
be used to analyze CMB polarization. Other possible applications include
stacked maps of cosmic structure, such as the galaxy
cluster catalog 2Mass \cite{2Mass}.\\

We finally mention that, although in this work we have focused on
testing isotropy while keeping within the Gaussian framework, 
our tools can also be used to search for deviations
from gaussianity in a completely model-independent way.

\subsection*{Acknowledgments}

We would like to thank Armando Bernui for useful suggestions and for
providing us with observational data in its final form. We also thank
the anonymous referee for clarifying that our test reduces
to the $\chi^2$ statistics, which led to
substantial improvements in the presentation of our work. 
This work was supported by Funda\c{c}\~ao de Amparo \`a 
Pesquisa do Estado de S\~ao Paulo (FAPESP), and by Brazil's Conselho 
Nacional de Desenvolvimento Cient\'{\i}fico e Tecnol\'ogico (CNPq.)

\appendix

\section*{Appendix A\label{appendixA}}

\subsection{Derivation of (\ref{Clmell})}

We will present here the details of the derivation of expression (\ref{Clmell}).
We start by equating expressions (\ref{fc-aniso}) and (\ref{full-cf})
\begin{equation}
\sum_{\ell}\sum_{l,m}\frac{2\ell+1}{\sqrt{4\pi}}\,\mathcal{C}_{\ell}^{lm}\, P_{\ell}(\cos\vartheta)
Y_{lm}(\hat{\mathbf{n}})=\sum_{\ell_{1},m_{1}}\sum_{\ell_{2},m_{2}}\langle 
a_{\ell_{1}m_{1}}a_{\ell_{2}m_{2}}^{*}\rangle Y_{\ell_{1}m_{1}}(\hat{\boldsymbol{n}}_{1})Y_{\ell_{2}m_{2}}(\hat{\boldsymbol{n}}_{2})\,.
\label{CLlm-alm}
\end{equation}
 As mentioned in the main text, the inversion of $\mathcal{C}_{\ell}^{lm}$
as a function of the $a_{\ell m}$'s is not a trivial task, since
the vector $\mathbf{n}$ depends non-linearly on the vectors $\hat{\boldsymbol{n}}_{1}$
and $\hat{\boldsymbol{n}}_{2}$. The easiest way to achieve this goal
is to pick up a coordinate system where only the $\vartheta$ dependence
(i.e., the modulus of the vector $\mathbf{n}$) is present. After
integrating it out, we rotate our coordinate system using three Euler
angles to recover back the $(\theta,\phi)$ dependence, which can
then be integrated with the help of some Wigner matrices identities.
We therefore start by positioning the vectors $\hat{\boldsymbol{n}}_{1}$
and $\hat{\boldsymbol{n}}_{2}$ in the $xy$ plane, i.e, we chose
$\hat{\boldsymbol{n}}_{1}=(\pi/2,\phi_{1})$, $\hat{\boldsymbol{n}}_{2}=(\pi/2,\phi_{2})$.
By (\ref{cos-theta}) we then have $\cos\vartheta=\cos(\phi_{1}-\phi_{2})$.
Using the relation \citep{AREdmonds}

\begin{equation}
Y_{\ell m}(\pi/2,\phi)=\lambda_{\ell m}e^{im\phi},\quad\mbox{where}\quad\lambda_{\ell m}
=\begin{cases}
(-1)^{\frac{\ell+m}{2}}\sqrt{\frac{2\ell+1}{4\pi}\frac{(\ell+m-1)!!}{(\ell+m)!!}\frac{(\ell-m-1)!!}{(\ell-m)!!}}\quad & 
\mbox{if}\quad\ell+m\in2\mathbb{N}\\
0 & \mbox{otherwise}
\end{cases}
\label{lambda-lm}
\end{equation}
 and $\mathbf{n}=(\sin\vartheta,0,0)$, we can integrate the $\vartheta$
dependence on both sides of (\ref{CLlm-alm}). This gives us 
\begin{equation}
\frac{1}{\sqrt{\pi}}\sum_{l,m}\mathcal{C}_{\ell}^{lm}Y_{lm}(0,0)=\sum_{\ell_{1},m_{1}}\sum_{\ell_{2},m_{2}}\langle 
a_{\ell_{1}m_{1}}a_{\ell_{2}m_{2}}^{*}\rangle I_{\ell_{1}m_{1}\ell_{2}m_{2}}^{\ell}
\label{CLlm-alm2}
\end{equation}
 where we have introduced the following definition 
\begin{equation}
I_{\ell_{1}m_{1}\ell_{2}m_{2}}^{\ell}\equiv-\lambda_{\ell_{1}m_{1}}\lambda_{\ell_{2}m_{2}}
\int_{0}^{\pi}P_{\ell}(\cos(\phi_{1}-\phi_{2}))e^{i(m_{1}\phi_{1}-m_{2}\phi_{2})}d(\cos(\phi_{1}-\phi_{2}))\,.
\label{I-ell-l1-l2-m}
\end{equation}

We need now to integrate out the $\theta$ and $\phi$ dependence
in the right-hand side of (\ref{CLlm-alm2}) which was hidden due
to our choice of a particular coordinate system. In order to do that,
we keep the vectors $\hat{\boldsymbol{n}}_{1}$ and $\hat{\boldsymbol{n}}_{2}$
fixed and make a rotation of our coordinate system using three Euler
angles $\omega=\{\alpha,\beta,\gamma\}$. This rotation changes the
coefficients $\mathcal{C}_{\ell}^{lm}$'s and $a_{\ell m}$'s according
to 
\[a_{\ell m}=\sum_{m^{\prime}}D_{mm^{\prime}}^{\ell}(\omega)\widetilde{a}_{\ell m^{\prime}}\,,
\qquad\mathcal{C}_{\ell}^{lm}=\sum_{m^{\prime}}D_{mm'}^{l}(\omega)\widetilde{\mathcal{C}}_{\ell}^{lm'}\]
 where $\widetilde{\mathcal{C}}_{\ell}^{lm}$ and $\widetilde{a}_{lm}^{}$
are the multipolar coefficients in the new coordinate system and where
$D_{mm'}^{l}(\omega)$ are the elements of the Wigner rotation matrix.
The advantage of positioning the vectors $\hat{\boldsymbol{n}}_{1}$
and $\hat{\boldsymbol{n}}_{2}$ in the plane $xy$ is that now the
angles $\theta$ and $\phi$ are given precisely by the Euler angles
$\beta$ and $\gamma$, regardless of the value of $\alpha$\[
\sum_{l,m}\mathcal{C}_{\ell}^{lm}Y_{lm}(0,0)=\sum_{l,m'}\widetilde{\mathcal{C}}_{\ell}^{lm'}
\left(\sum_{m}D_{mm'}^{l}(\alpha,\beta,\gamma)Y_{lm}(0,0)\right)=\sum_{l,m'}\widetilde{\mathcal{C}}_{\ell}^{lm'}Y_{l,-m}(\beta,\gamma)\]
 where in the last step we have used $Y_{lm}(0,0)=\sqrt{(2l+1)/4\pi}\,\delta_{m0}$.
Therefore, in our new coordinate system we have (dropping the {}``
$\widetilde{}$ '' in our notation) \[
\frac{1}{2\pi}\sum_{l,m}\mathcal{C}_{\ell}^{lm}D_{0m}^{l}(\omega)\sqrt{2l+1}=\sum_{\ell_{1},m_{1}}\sum_{\ell_{2},m_{2}}\langle a_{\ell_{1}m_{1}}a_{\ell_{2}m_{2}}^{*}\rangle\sum_{m_{1}^{\prime}m_{2}^{\prime}}I_{\ell_{1}m_{1}^{\prime}\ell_{2}m_{2}^{\prime}}^{\ell}D_{m_{1}^{\prime}m_{1}}^{\ell_{1}}(\omega)D_{m_{2}^{\prime}m_{2}}^{\ell_{2}*}(\omega)\,.\]
 We may now isolate $\mathcal{C}_{\ell}^{lm}$ using the identities
\citep{AREdmonds} 
\begin{eqnarray*}
\int d\omega\, D_{m_{1}m_{1}^{\prime}}^{l_{1}*}(\omega)D_{m_{2}m_{2}^{\prime}}^{l_{2}}(\omega) & = & \frac{8\pi^{2}}{2l_{1}+1}\delta_{l_{1}l_{2}}\delta_{m_{1}m_{2}}\delta_{m_{1}^{\prime}m_{2}^{\prime}}\\
\int d\omega\, D_{m_{1}^{\prime}m_{1}}^{l_{1}}(\omega)D_{m_{2}^{\prime}m_{2}}^{l_{2}}(\omega)D_{m_{3}^{\prime}m_{3}}^{l_{3}}(\omega) & = & 8\pi^{2}\left(\begin{array}{ccc}
l_{1} & l_{2} & l_{3}\\
m_{1}^{\prime} & m_{2}^{\prime} & m_{3}^{\prime}\end{array}\right)\left(\begin{array}{ccc}
l_{1} & l_{2} & l_{3}\\
m_{1} & m_{2} & m_{3}\end{array}\right)
\end{eqnarray*}
 where $d\omega=\sin\beta d\beta d\alpha d\gamma$, to obtain 
\[\frac{1}{\sqrt{2l+1}}\mathcal{C}_{\ell}^{lm}=2\pi\sum_{\ell_{1},m_{1}}\sum_{\ell_{2},m_{2}}\langle a_{\ell_{1}m_{1}}a_{\ell_{2}m_{2}}^{*}\rangle\sum_{m_{1}^{\prime}m_{2}^{\prime}}I_{\ell_{1}m_{1}^{\prime}
\ell_{2}m_{2}^{\prime}}^{\ell}(-1)^{m_{2}+m_{2}^{\prime}+m}\left(\begin{array}{ccc}
\ell_{1} & \ell_{2} & l\\
m_{1}^{\prime} & -m_{2}^{\prime} & 0\end{array}\right)\!\left(\begin{array}{ccc}
\ell_{1} & \ell_{2} & l\\
m_{1} & -m_{2} & -m\end{array}\right)\,.\]
 If we now do the redefinitions
\[-m_{2}\rightarrow m_{2}\,,\quad-m\rightarrow m\,,\quad\mathcal{C}_{\ell}^{lm}
\rightarrow(-1)^{m}\mathcal{C}_{\ell}^{l,-m}\]
 and note that the first 3J symbol above is identically zero unless
$m'_{1}=m'_{2}$, we obtain finally (\ref{Clmell}).

\subsection{Useful identities}

We present here some useful identities related to the 3J symbols: 
\begin{itemize}
\item Isotropic limit \[
\left(\begin{array}{ccc}
l_{1} & l_{2} & 0\\
m_{1} & m_{2} & 0\end{array}\right)=\frac{(-1)^{l_{2}-m_{1}}}{\sqrt{2l_{1}+1}}\delta_{l_{1}l_{2}}\delta_{m_{1},-m_{2}}\]

\item Parity and permutations 
\begin{eqnarray*}
\left(\begin{array}{ccc}
l_{1} & l_{2} & l\\
m_{1} & m_{2} & m\end{array}\right) & = & \left(\begin{array}{ccc}
l & l_{1} & l_{2}\\
m & m_{1} & m_{2}\end{array}\right)\\
 & = & (-1)^{l_{1}+l_{2}+l}\left(\begin{array}{ccc}
l_{2} & l_{1} & l\\
m_{2} & m_{1} & m\end{array}\right)\\
 & = & (-1)^{l_{1}+l_{2}+l}\left(\begin{array}{ccc}
l_{1} & l_{2} & l\\
-m_{1} & -m_{2} & -m\end{array}\right)
\end{eqnarray*}

\item Orthogonality 
\begin{eqnarray*}
\sum_{m_{1}=-l_{1}}^{l_{1}}\sum_{m_{2}=-l_{2}}^{l_{2}}\left(\begin{array}{ccc}
l_{1} & l_{2} & l_{3}\\
m_{1} & m_{2} & m_{3}\end{array}\right)\left(\begin{array}{ccc}
l_{1} & l_{2} & l_{3}^{\prime}\\
m_{1} & m_{2} & m_{3}^{\prime}\end{array}\right) & = & \frac{\delta_{l_{3}l_{3}^{\prime}}\delta_{m_{3}m_{3}^{\prime}}}{2l_{3}+1}\\
\sum_{l_{1}=|l_{2}-l_{3}|}^{l_{2}+l_{3}}\sum_{m_{1}=-l_{1}}^{l_{1}}(2l+1)\left(\begin{array}{ccc}
l_{1} & l_{2} & l_{3}\\
m_{1} & m_{2} & m_{3}\end{array}\right)\left(\begin{array}{ccc}
l_{1} & l_{2} & l_{4}\\
m_{1} & m_{2}^{\prime} & m_{3}^{\prime}\end{array}\right) & = & \delta_{m_{2}m_{2}^{\prime}}\delta_{m_{3}m_{3}^{\prime}}\\
\sum_{m=-l}^{l}(-1)^{l-m}\left(\begin{array}{ccc}
l & l & \ell\\
m & -m & 0\end{array}\right) & = & \sqrt{2l+1}\delta_{\ell,0}\,.
\end{eqnarray*}
 The last expression is particularly useful in the derivation of (\ref{SI}). 
\end{itemize}

\subsection{Some properties of the integral (\ref{Int-l-ell})}

The geometrical coefficients $I_{\ell_{1}\ell_{2}}^{l,\ell}$ defined
in (\ref{Int-l-ell}) has many interesting properties which can be
explored in order to speed up numerical computation of (\ref{Clmell}).
First, we note that it is symmetric under permutation of $\ell_{1}$
and $\ell_{2}$

\begin{eqnarray*}
I_{\ell_{1}\ell_{2}}^{l,\ell} & = & \sum_{m}I_{\ell_{1}m\ell_{2}m}^{\ell}(-1)^{m}\left(\begin{array}{ccc}
\ell_{1} & \ell_{2} & l\\
m & -m & 0\end{array}\right)\\
 & = & \sum_{m}I_{\ell_{2}m\ell_{1}m}^{\ell}(-1)^{m+\ell_{1}+\ell_{2}+l}\left(\begin{array}{ccc}
\ell_{2} & \ell_{1} & l\\
-m & m & 0\end{array}\right)\\
 & = & \sum_{m}I_{\ell_{2}m\ell_{1}m}^{\ell}(-1)^{m+2(\ell_{1}+\ell_{2}+l)}\left(\begin{array}{ccc}
\ell_{2} & \ell_{1} & l\\
m & -m & 0\end{array}\right)\\
 & = & I_{\ell_{2}\ell_{1}}^{l,\ell}\,.
\end{eqnarray*}
 Some of the other properties are a consequence of the integral $I_{\ell_{1}m\ell_{2}m}^{\ell}$
defined in (\ref{I-ell-l1-l2-m}). We may note for example that, due
to the symmetry of the $\lambda_{\ell m}$ coefficient defined in
(\ref{lambda-lm}), we will have

\[I_{\ell_{1}\ell_{2}}^{l,\ell}=0\,,\quad\mbox{for any}\quad\{(\ell_{1},\ell_{2})\in\mathbb{N}\,|\,\ell_{1}+\ell_{2}=\mbox{odd}\}\,.\]

Furthermore, the $\lambda_{\ell m}$ coefficients restrict the $m$
summation above to their values which obey $m+\ell_{1}+\ell_{2}=\mbox{even}$.
If we further notice that (\ref{I-ell-l1-l2-m}) is proportional to
the integral of a integral of the form $\int_{-1}^{1}P_{\ell}(\cos\theta)\cos m\theta\, d\theta$,
and that this integral is zero unless $\ell+m=\mbox{even}$, we conclude
that 
\[I_{\ell_{1}\ell_{2}}^{l,\ell}=0\,,\quad\mbox{for any}\quad\{(\ell_{1},\ell_{2},\ell)\in\mathbb{N}\,|\,\ell_{1}+\ell_{2}+\ell=\mbox{odd}\}\]

Besides, using the fact that the integral $\int_{-1}^{1}P_{\ell}(\cos\theta)\cos m\theta\, d\theta$
is zero for any $m<\ell$, we find

\[
I_{\ell_{1}\ell_{2}}^{l,\ell}=0\,,\quad\mbox{for any}\quad\{(\ell_{1},\ell_{2},\ell)\in\mathbb{N}\,|\,\ell_{1}<\ell,\ell_{2}<\ell\}.\]

We finally comment on the special case where $l=0$, for which we
have

\[
I_{\ell_{1}\ell_{2}}^{0,\ell}=\frac{(-1)^{\ell_{1}}}{\sqrt{2\ell_{1}+1}}\left(\sum_{m}I_{\ell_{1}m\ell_{1}m}^{\ell}\right)\delta_{\ell_{1}\ell_{2}}\,.\]
 However 
\begin{eqnarray*}
\sum_{m=-\ell'}^{\ell'}I_{\ell'm\ell'm}^{\ell} & = & \int_{0}^{\pi}P_{\ell}(\cos\vartheta)\left(\sum_{m=-\ell'}^{\ell'}\frac{(2\ell'+1)}{4\pi}\frac{(\ell'+m-1)!!}{(\ell'+m)!!}\frac{(\ell'-m-1)!!}{(\ell'-m)!!}e^{im\vartheta}\right)d(-\cos\vartheta),\quad\ell'+m=\mbox{even}\\
 & = & \frac{2\ell'+1}{4\pi}\int_{-1}^{1}P_{\ell}(x)P_{\ell'}(x)dx\\
 & = & \frac{1}{2\pi}\delta_{\ell\ell'}\,.
\end{eqnarray*}
 where in the derivation above we have made use of the Fourier series
expansion of the Legendre polynomial. So we conclude that 
\begin{equation}
I_{\ell_{1}\ell_{2}}^{0,\ell}=\frac{(-1)^{\ell_{1}}}{2\pi\sqrt{2\ell_{1}+1}}
\delta_{\ell\ell_{1}}\delta_{\ell_{1}\ell_{2}},\label{IL-zero}
\end{equation}
 which is needed in the derivation of (\ref{SI}) and (\ref{Hlell}).

\begin{eqnarray*}
\end{eqnarray*}

\bibliographystyle{h-physrev} \addcontentsline{toc}{section}{\refname}\bibliographystyle{h-physrev}
\bibliography{angular-planar}

\end{document}